# Resilience and Volatility in Academic Publishing:
## The Case of the University of Maribor (2004–2023)

Mojca Tancer Verboten (1,2), Dean Korošak (3,4)*

(1) University of Maribor, Faculty of Law; mojca.tancer@um.si
(2) University of Maribor, Faculty of Chemistry and Chemical Engineering
(3) University of Maribor, Faculty of Civil Engineering, Transportational Engineering and Architecture; dean.korosak@um.si
(4) University of Maribor, Faculty of Medicine; dean.korosak@um.si
*Correspondence: dean.korosak@um.si

**Abstract:** This article examines the resilience and volatility of academic publishing at the University of Maribor (UM) from 2004 to 2023, a period marked by significant economic and policy shifts in Slovenia mainly through changes to higher education legislation and a new way of financing universities. Using employment data from UM's internal records and publication data from OpenAlex, we analyze the relationship between the number of employed researchers and the number of unique authors publishing under the UM affiliation. Despite a significant drop in researcher employment during the economic recession and austerity period (2009-2013), the number of unique authors publishing under the UM affiliation surprisingly increased, driven by factors such as a shift to project-based funding, contributions from a growing doctoral student cohort, and increasing international collaborations. Analysis of author turnover, using both annual and 5-year cumulative windows, reveals a striking dichotomy: high short-term volatility (annual Jaccard Index ~0.32-0.38, churn ~40-50%) contrasted with significant mid-term stability (5-year Jaccard Index ~0.75-0.82, churn ~8-10%). Survival analysis, employing both cohort-based and Kaplan-Meier methods, confirms this pattern, revealing high initial attrition among publishing authors followed by long-term persistence for a core group of researchers. Furthermore, network analysis of co-authorship patterns reveals that the UM research network has become increasingly resilient to the targeted removal of influential authors over time. Most significantly, we identify a fundamental shift in network structure around 2016, when the co-authorship network transitioned from dissassortative to assortative mixing patterns, indicating a profound change in collaboration dynamics. We discuss the implications of these findings for research policy and university management, emphasizing the need to balance short-term performance metrics with the long-term stability and resilience of the research community.

Keywords: Academic Publishing, Researcher Mobility, Network Analysis, Survival Analysis, Research Policy

## 1. Introduction

Universities worldwide serve as cornerstones of knowledge creation, their scientific output driven by a complex interplay of human capital (researchers, faculty, and staff), research funding (both internal and external), and institutional infrastructure (Auranen & Nieminen, 2010; Geuna & Martin, 2003). This intricate ecosystem is constantly evolving, influenced by national and international policies, economic conditions, and evolving research practices. The Slovenian academic landscape, having undergone significant transformations over the past two decades, provides a particularly compelling case study for examining the resilience and volatility of academic publishing in response to economic and policy changes (Mali, 2013).

Several key factors have shaped the Slovenian research environment during the period from 2004 to 2023. In 2004, a comprehensive amendment to higher education legislation was adopted, namely Zakon

o spremembah in dopolnitvah Zakona o visokem šolstvu (ZViS-D, Act on Amendments to the Higher Education Act) and the Regulation on Public Financing of Higher Education and Other Institutions, Member Universities, from 2004 to 2008 (Regulation), which changed the system of financing public universities in Slovenia, including the UM. The global economic recession of the late 2000s and the subsequent Eurozone crisis led to significant austerity measures, including the 2012 Zakon za uravnoteženje javnih financ (ZUJF; Act for Balancing Public Finances) in Slovenia. This legislation imposed substantial cuts on public spending, directly impacting university budgets and research funding (Cadez et al., 2017). Concurrently, a growing emphasis on competitive, project-based funding, from both national (e.g., the Slovenian Research and Innovation Agency, ARIS) and international sources (e.g., the European Union's Horizon programs), has created a more dynamic, but potentially more precarious, research environment (Hicks, 2012).

Another critical factor is the expansion of doctoral and postdoctoral programs in Slovenia. This expansion has increased the pool of early-career researchers actively contributing to scientific output, often through short-term contracts or project-based affiliations (Kyvik & Aksnes, 2015). This influx of researchers, while beneficial for overall productivity, can also contribute to higher turnover rates and fluctuations in authorship patterns.

Arguably the most pervasive influence on the Slovenian research landscape has been the increasing emphasis on metric-based evaluation of research performance, using bibliometric indicators such as publication counts, citations, h-index, and journal impact factors (Franzoni et al., 2011). It is necessary to highlight two more extensive changes to the internal acts at UM that determine the procedure for election to a title. In 2004, UM adopted the Criteria for Election to the Titles of Higher Education Teachers, Scientific Workers and Higher Education Associates, which precisely define what is taken into account when assessing scientific research activity when electing to a title. The next major change to internal rules at UM occurred in 2012, with the adoption of new Criteria for Appointment to Titles, based on the Minimum Standards for Appointment to Titles, which were adopted at the national level by the National Agency for Quality Assurance in Higher Education (NAKVIS). In 2011, the Research and Innovation Strategy of Slovenia 2011-2020 was adopted for the first time. The Slovenian Current Research Information System (SICRIS) serves as a central repository for tracking researchers, projects, and publications, making these metrics readily available and highly influential. This emphasis on quantifiable output has incentivized researchers to prioritize publication in high-impact journals, collaborate internationally, and register all their publications, potentially altering traditional publishing behaviors and the composition of the researcher community (Hicks, 2012).

Previous studies in other national contexts have suggested that even during periods of fiscal constraint and reduced direct employment, the broader population of active researchers can continue to grow, especially if doctoral students, guest researchers, and short-term project collaborators remain engaged in publishing through co-authorship (Abramo et al., 2017; Lee & Bozeman, 2005). However, the long-term consequences of these trends on the stability and resilience of the research ecosystem remain an area of ongoing investigation. Specifically, it's crucial to understand not just the number of researchers, but also the dynamics of their engagement—how stable or volatile the research workforce is over time (Kwiek & Szymula, 2024; Levin & Stephan, 1991).

This paper investigates these dynamics at the University of Maribor (UM), the second-largest university in Slovenia, over the period 2004-2023. We examine the evolution of researcher employment alongside two key measures of scientific output derived from the OpenAlex bibliographic database (Priem et al., 2022): the number of unique authors affiliated with UM and the overlap of these authors across different time windows (annual and cumulative 5-year periods). We also investigate the evolution of the structure of the co-authorship network using methods established in the network science literature (Barabási et al., 2002; Newman, 2001).

By combining quantitative analysis of employment and publication data with network analysis of co-authorship patterns, we address the following central research question: how has the interplay between researcher employment, scientific output (quantified by unique authorship in the OpenAlex database), and co-authorship network structure at the University of Maribor evolved from 2004 to 2023—a period characterized by economic fluctuations, policy changes, and an increasing emphasis on metric-based evaluation—and what do these dynamics reveal about the resilience and volatility of the university's research ecosystem? By addressing this question, we aim to contribute to a deeper understanding of the factors that influence the stability and productivity of research institutions in a changing global landscape, offering insights relevant to research policy and university management.

## 2. Materials and Methods

This study employs a mixed-methods approach, combining quantitative analysis of employment and publication data with network analysis of co-authorship patterns. We analyze data from the University of Maribor (UM) spanning the period from 2004 to 2023. For our primary data sources, we utilized internal human resources records from UM and the OpenAlex bibliographic database. The combination of these sources allows us to link employment status with publication activity, providing a comprehensive view of the research landscape at UM. Employment data consisted of the total number of employed researchers and teaching faculty with research responsibilities for each year from 2004 to 2023, provided in an aggregated, anonymized format. For publication data, we selected OpenAlex due to its comprehensive coverage, open access nature, unique author identifiers, and rich metadata (Priem et al., 2022). We queried the database for all publications listing "University of Maribor" as an author affiliation in the study period.

To analyze both short-term and longer-term trends in authorship, we defined two types of time windows: annual sets (containing unique authors who published with UM in a given calendar year) and five-year cumulative windows (combining authors from consecutive five-year intervals). To quantify the stability and turnover in these sets, we calculated two key metrics: the Jaccard Index (measuring similarity between sets) and the churn rate (quantifying the proportion of authors lost from one period to the next).

For analyzing the longevity of authors' publishing careers at UM, we employed two complementary survival analysis techniques: a cohort-based approach and the Kaplan-Meier estimator. The cohort-based analysis tracked the survival fraction of publishing authors based on their first year of publication with a UM affiliation. The Kaplan-Meier estimator, which handles right-censored data, provided a more statistically robust measure of long-term survival patterns.

To investigate the structure and resilience of the co-authorship network, we constructed networks for different time windows and analyzed their properties. We calculated various network metrics including the Largest Connected Component (LCC), betweenness centrality, clustering coefficient, and assortativity. To assess the resilience of the network to the removal of key authors, we performed targeted attacks by iteratively removing nodes in descending order of their betweenness centrality, following established methodologies (Albert et al., 2000; Callaway et al., 2000). We compared the resilience of the original co-authorship network to that of a degree-preserving rewired network to isolate the effect of network structure on resilience.

Additionally, we applied community detection algorithms to identify groups of authors who are more densely connected to each other than to authors in other groups, measuring the similarity of community structures across different time windows using Normalized Mutual Information (NMI) and Adjusted Rand Index (ARI).

For data quality assurance, we crosschecked the OpenAlex database with Web of Science/InCites data, finding an overlap of over 85% for records containing DOIs. All data processing, analysis, and visualization were performed using Python, with the code publicly available to promote transparency and reproducibility.

## 3. Results

Over 2004-2007, UM saw a notable increase in employed researchers (from ~1100 to 1200+), it should be noted that in 2003, UM established new faculties and expanded its research areas to include medicine, logistics, criminology and energy technology, followed by a sharp drop around 2009-2013—coinciding with the global financial crisis and austerity measures under ZUJF. This dip bottomed out near 1000 employees before recovering and surpassing pre-crisis levels around 2018-2020. In contrast, the number of unique authors affiliated with UM showed a steady or even accelerating increase from ~300-400 in the early years (2004-2006) to over 1000 by 2020-2022. Notably, during the 2009-2013 period, when employment dropped, the total number of unique authors continued to climb.

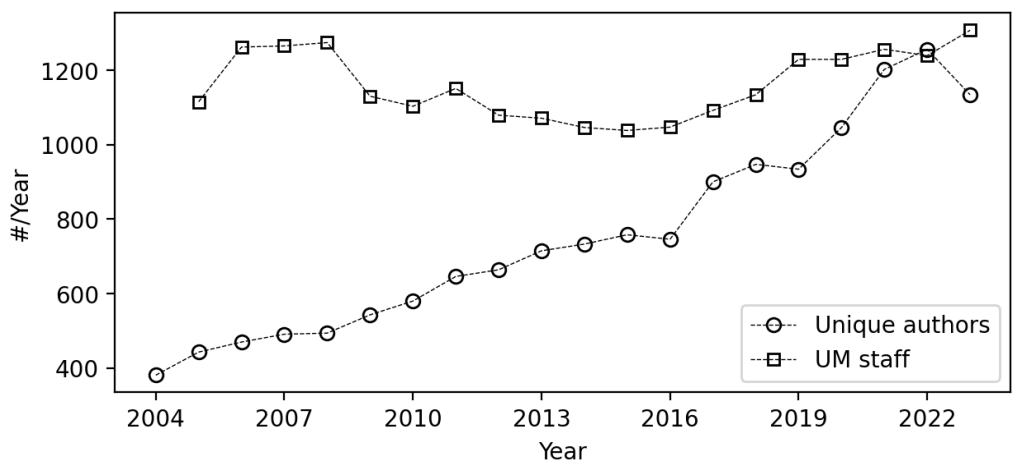

**Figure 1. Unique Evolution of Unique Authors and Research Staff at the University of Maribor (2004-2023).** Number of unique authors per year (circles) and number of teaching and full-time research staff (squares) at the University of Maribor from 2004 to 2023. Note the divergence between 2011 and 2016, when staff numbers declined while the count of unique authors continued to increase, indicating sustained research output despite workforce reductions.

Figure 1 illustrates these divergent trends. While staff numbers declined between 2011 and 2016, the count of unique authors increased, indicating that factors beyond sheer staff numbers contributed to sustained or enhanced productivity. This pattern aligns with findings from other academic systems where research output has continued to grow despite resource constraints (Abramo et al., 2017; Auranen & Nieminen, 2010).

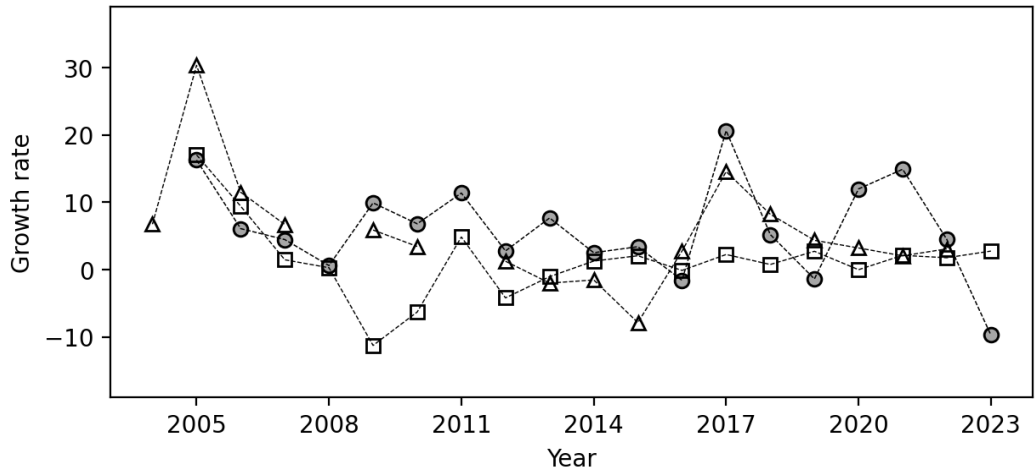

**Figure 2. Comparative Growth Rates of Researchers in Slovenia (2004-2023).** Annual growth rates of unique authors affiliated with the University of Maribor (circles), UM teaching and full-time research staff (squares), and OECD data for growth rate of researchers in Slovenia (triangles). The graph highlights different magnitudes and timing of responses to economic cycles and policy changes, with unique authors maintaining positive growth even during periods of staff reduction.

Figure 2 compares the annual growth rates of three overlapping populations: (1) unique authors who published with UM affiliations, (2) UM teaching and full-time research staff, and (3) OECD data on researchers in Slovenia. During the economic downturn (2011-2016), UM staff growth declined significantly while unique author growth remained consistently positive, peaking at ~30% around 2013. Slovenia's overall researcher growth shows milder fluctuations than UM's staff numbers. Post-2016, all three metrics exhibit general alignment in their growth patterns. These divergent trends suggest complex adaptation mechanisms in the academic system, as noted by Franzoni et al. (2011), who found that increased emphasis on publication metrics can drive behavioral changes among researchers.

Our analysis of one-year overlap reveals high short-term volatility in the author population. Calculating the Jaccard index from one year to the next yields values mostly in the range of 0.32-0.38, indicating that only around one-third of each year's authors also publish the following year. Annual churn hovers around 40-50%, suggesting considerable short-term flux—many authors appear only once or intermittently. This pattern is consistent with findings from larger-scale studies of scientific careers (Kwiek & Szymula, 2024).

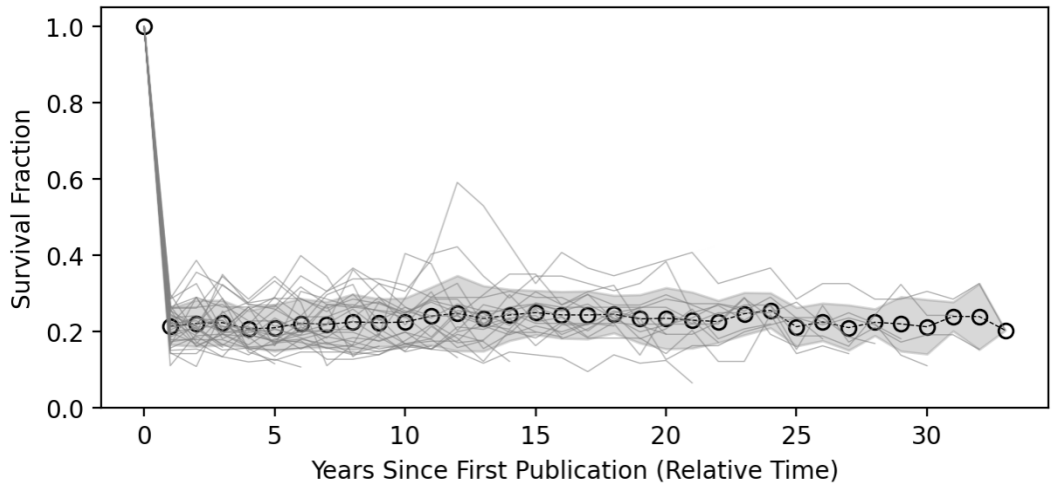

**Figure 3. Survival Fraction of Publishing Authors at the University of Maribor.** Cohort-based survival analysis showing the fraction of authors continuing to publish as a function of years since first publication. The mean survival fraction (circles) shows a rapid initial decline followed by stabilization around 20%. Individual cohorts (thin grey lines) exhibit considerable variability, with the grey shaded area representing standard deviation, demonstrating intermittent publishing patterns across career trajectories.

Figure 3 shows the survival fraction of publishing authors at UM. The graph reveals a very steep drop in the survival fraction in the first few years after an author's first publication, with the average survival fraction dropping to around 20% within 5 years. After this initial sharp decline, the survival fraction stabilizes, suggesting that a core group of authors remains active in publishing over the long term. These patterns align with broader trends in academic career trajectories documented by Levin and Stephan (1991) and Kwiek and Szymula (2024).

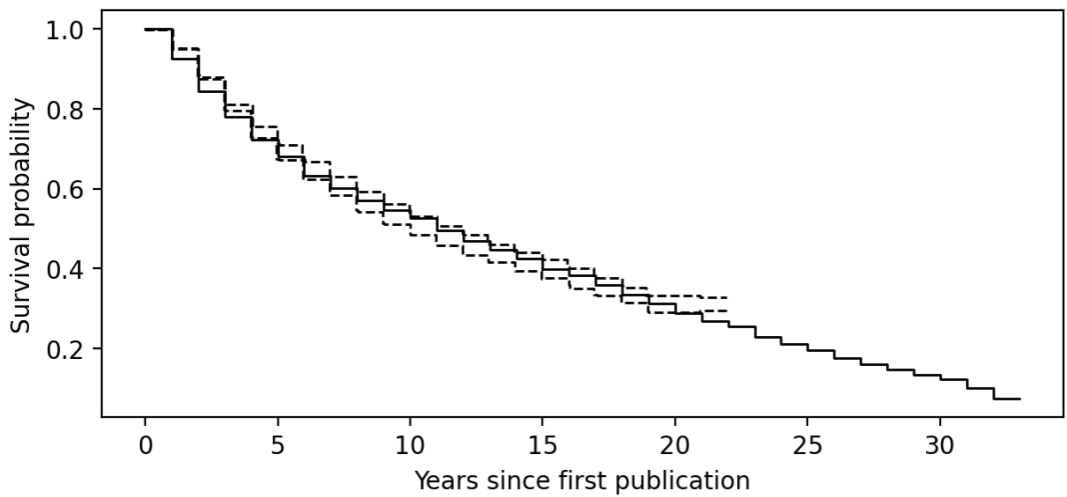

**Figure 4. Kaplan-Meier Survival Analysis of Publishing Authors.** Survival probability estimated by the Kaplan-Meier method (solid line), showing the likelihood of continued publishing activity over time since first publication. This statistically robust measure accounts for right-censoring of authors who may still be active at the end of the observation period. Comparative data from Kwiek & Szymula (2024) on attrition in science overall is referenced for context (dashed lines).

Figure 4 presents the Kaplan-Meier estimate of survival probability, which correctly accounts for right-censoring. This analysis suggests higher long-term survival than raw cohort averages, with approximately 20% of authors estimated to still be publishing 20+ years after their first publication.

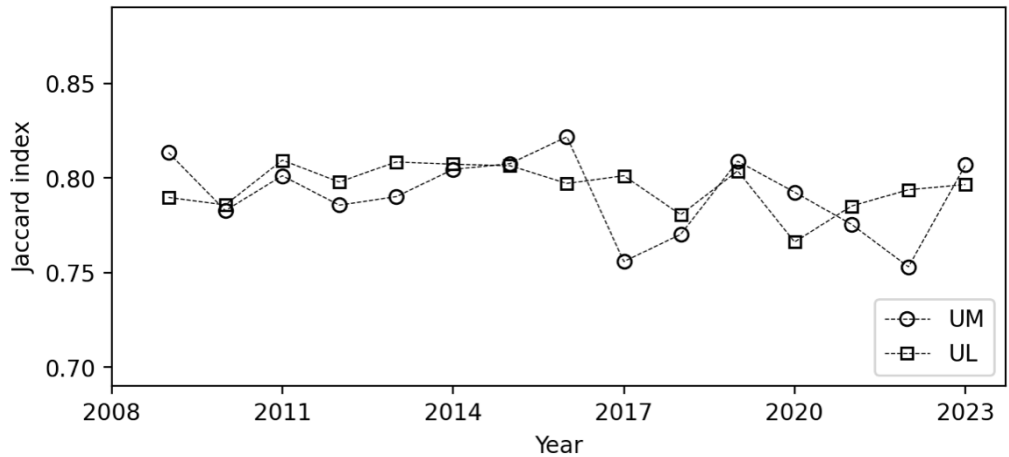

**Figure 5. Five-Year Author Stability at Two Slovenian Universities.** Jaccard index calculated over consecutive 5-year windows for the University of Maribor (circles, dashed line) and University of Ljubljana (squares, dashed line) from 2008 to 2023. The high values (0.75-0.85) indicate substantial overlap between authors publishing in consecutive 5-year periods at both institutions, demonstrating similar patterns of mid-term stability.
.

When we switch to 5-year windows, both Jaccard and churn metrics change drastically. The 5-year Jaccard index ranges around 0.75-0.82, implying that the majority of authors in a given 5-year interval also appear in the next 5-year interval. The 5-year churn drops to ~7-11%, far lower than the one-year churn. This smoother perspective reveals a substantial "core" of authors who remain active on a multi-year basis, despite the apparent short-term volatility.

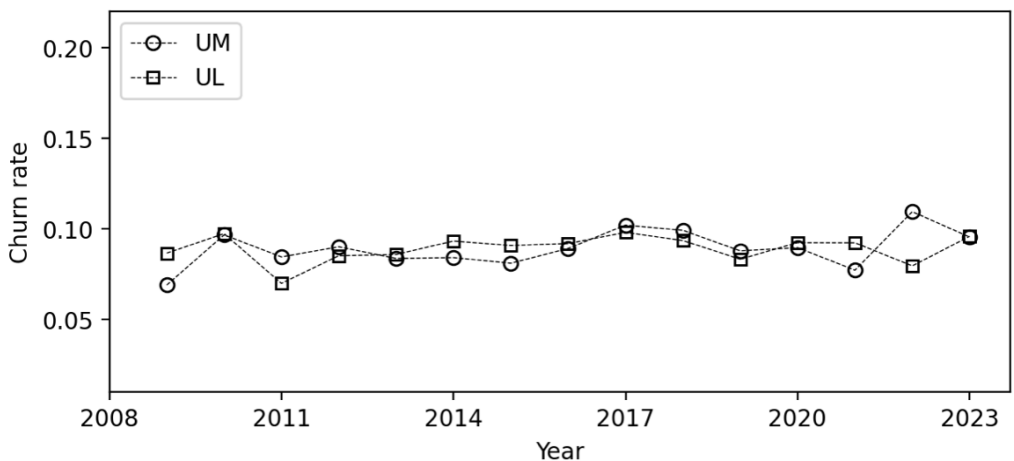

**Figure 6. Author Turnover at Two Slovenian Universities.** Churn rate calculated over consecutive 5-year windows for the University of Maribor (circles, dashed line) and University of Ljubljana (squares, dashed line) from 2008 to 2023. The consistently low values (5-10%) confirm high author retention at both universities, complementing the Jaccard index findings.

Figures 5 and 6 compare the 5-year Jaccard indices and churn rates between the University of Maribor and the University of Ljubljana. Both universities show relatively high Jaccard indices (0.75-0.85) and low churn rates (5-10%), indicating substantial author stability over 5-year periods. The patterns are remarkably similar between the two institutions, suggesting that these dynamics may reflect broader national trends rather than institution-specific factors.

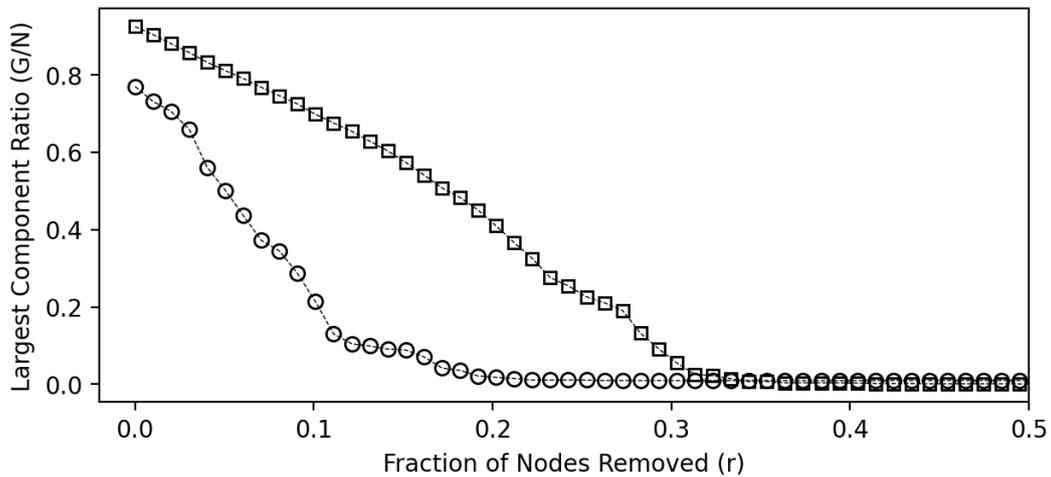

**Figure 7. Fragility Network Fragility Under Targeted Node Removal.** Resilience analysis showing the ratio of the largest connected component (G/N) as a function of the fraction of nodes removed (r) under targeted attacks based on betweenness centrality. The original co-authorship network (circles) fragments more rapidly than its degree-preserving randomly rewired counterpart (squares), indicating specific vulnerabilities in the actual collaboration structure.

Network analysis of co-authorship patterns provides additional insights. Figure 7 shows the largest connected component (LCC) ratio as a function of the fraction of nodes removed under targeted attacks. The original network is significantly more vulnerable to targeted attacks than a degree-preserving randomly rewired version, with the LCC ratio dropping much faster. This vulnerability aligns with findings from other complex networks (Albert et al., 2000; Callaway et al., 2000).

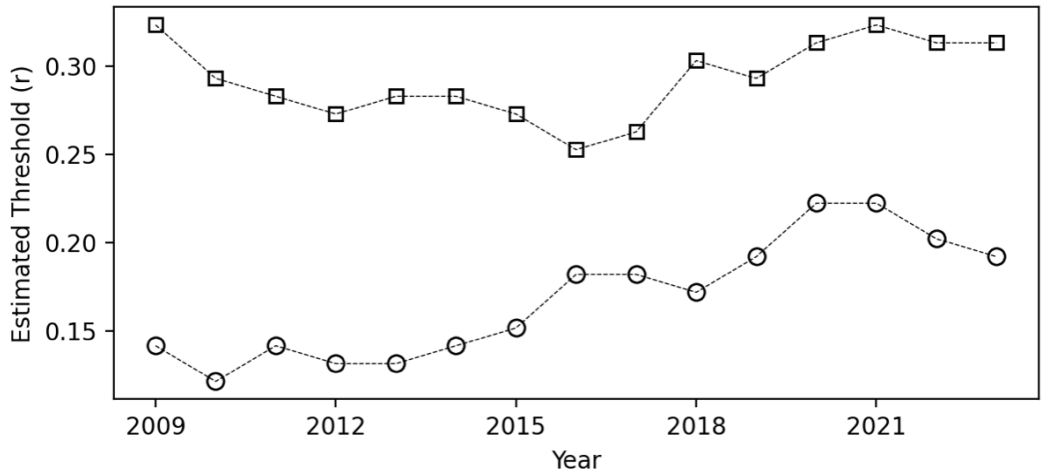

**Figure 8. Evolution of Network Resilience Over Time.** Threshold fraction of nodes that must be removed to cause network collapse (estimated threshold r) for each year from 2009 to 2023. The original co-authorship network (circles) shows a gradual increase in this threshold over time, indicating growing resilience to targeted attacks, while remaining consistently more vulnerable than its rewired counterpart (squares).

Figure 8 tracks how this vulnerability has changed over time. The threshold for network collapse has gradually increased for the UM network, suggesting that the co-authorship structure has become more robust to targeted attacks in recent years. This increasing resilience could be attributed to network growth, increased interdisciplinary collaboration, and strategic university initiatives promoting research collaboration.

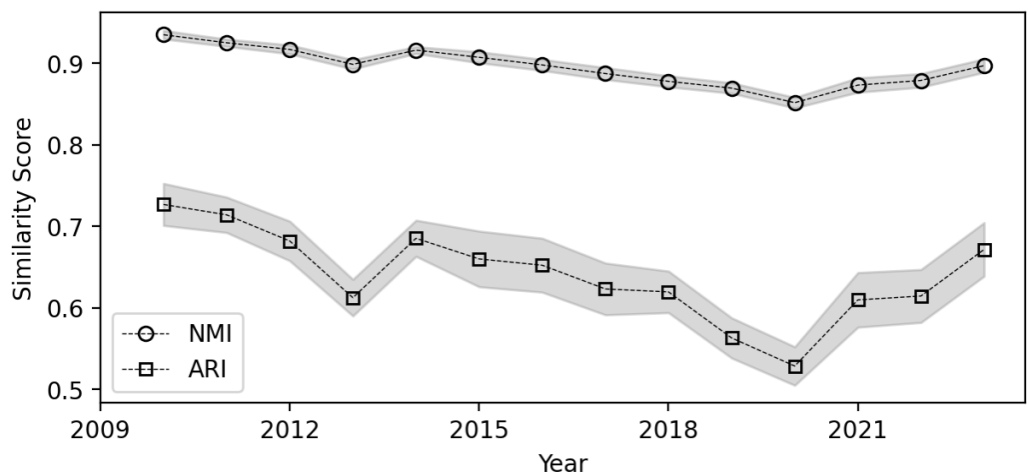

**Figure 9. Persistence Stability of Research Community Structure.** Similarity of community structures in co-authorship networks across consecutive 5-year windows, measured by Normalized Mutual Information (NMI, circles) and Adjusted Rand Index (ARI, squares). The consistently high NMI (~0.9) indicates stability in broad community structure, while lower ARI values show more variability in exact cluster memberships. The dip in ARI around 2016-2019 followed by recovery suggests a period of structural reorganization.

Figure 9 provides insight into how stable the co-authorship communities remain from year to year, using Normalized Mutual Information (NMI) and Adjusted Rand Index (ARI). The NMI curve is consistently high—hovering around 0.9—indicating that the broad community structure does not radically change over time. In contrast, the ARI curve is lower and shows more pronounced fluctuations, suggesting that on a finer-grained level, the communities shift more noticeably. The dip in ARI around 2016-2019, followed by a recent uptick, points to a period of rapid structural reorganization in the co-authorship network, followed by partial reconsolidation—timing that aligns with the observed shift in assortativity.

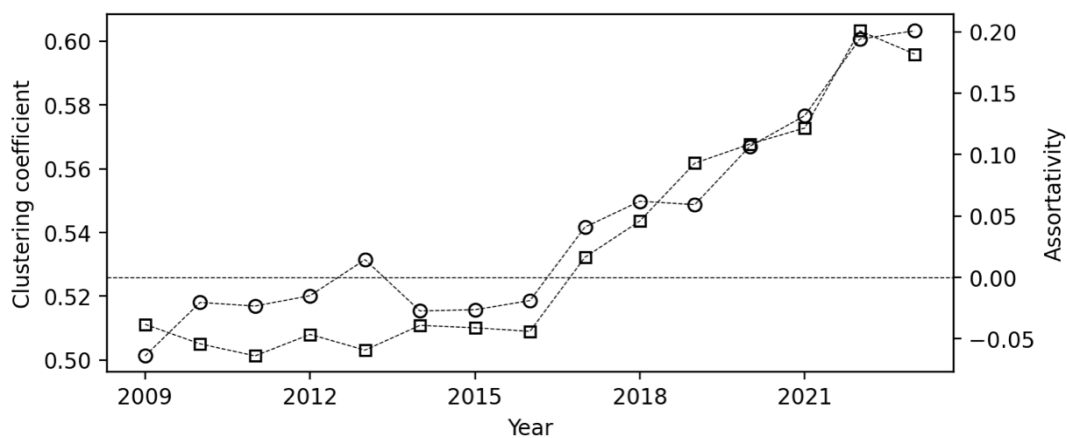

**Figure 10. Evolution of Network Assortativity and Clustering.** Clustering coefficient and assortativity of the UM co-authorship network from 2009 to 2023. Clustering remains stable around 0.53 until 2015, then increases to 0.6 by 2023. Assortativity undergoes a dramatic shift from negative (dissassortative) to positive (assortative) values around 2016, marking a fundamental change in collaboration patterns from diverse connections to more stratified structures.

A particularly significant finding emerges in Figure 10, which shows the evolution of clustering coefficient and assortativity in the UM co-authorship network from 2009 to 2023. The clustering coefficient remained relatively stable around 0.53 from 2009 to 2015 before increasing to 0.6 by 2023. More dramatically, the network's assortativity underwent a fundamental shift, transitioning from negative values (dissassortative) to positive values (assortative) around 2016, and continuing to increase to 0.2 by 2023.

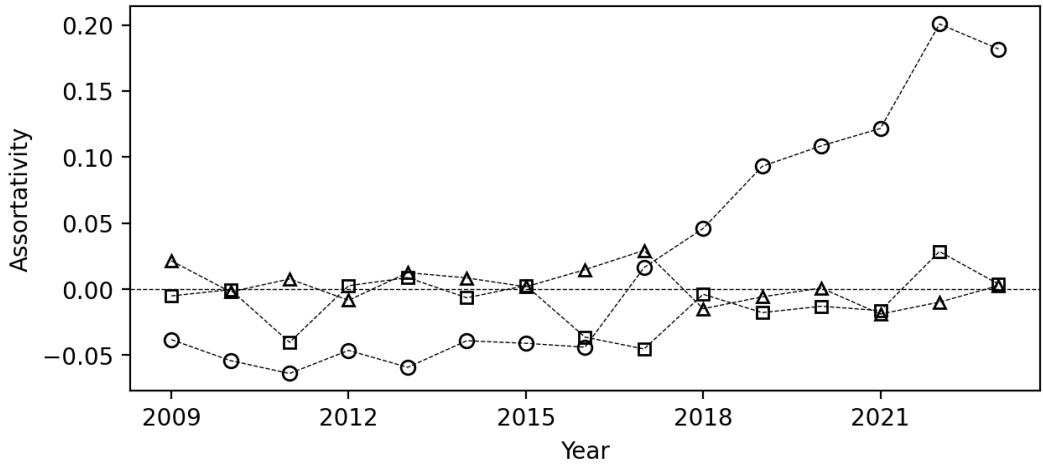

**Figure 11. Validation of Assortativity Shift as a Structural Phenomenon.** Comparison of assortativity in the original UM co-authorship network (circles) with two types of randomly rewired versions (squares and triangles) from 2009 to 2023. The randomly rewired networks maintain assortativity close to zero, confirming that the observed shift to positive assortativity represents a genuine structural change rather than a statistical artifact.

This shift in assortativity represents a profound change in collaboration patterns. In network science terms, a dissassortative network indicates that highly connected authors tend to collaborate with less connected authors, while an assortative network indicates that highly connected authors primarily collaborate with other highly connected authors (Newman, 2002). Figure 11 confirms that this shift is a genuine structural change rather than a statistical artifact, as randomly rewired networks maintain assortativity values close to zero.

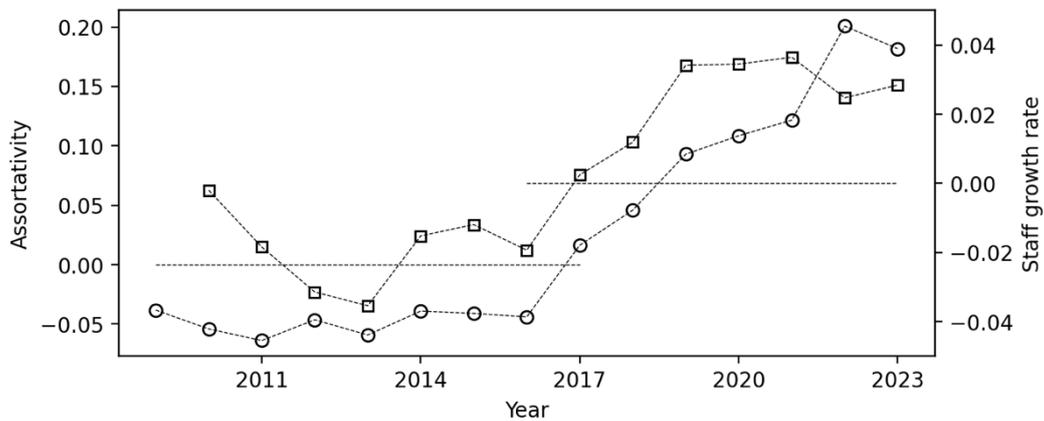

**Figure 12. Assortativity Parallel Evolution of Network Structure and Workforce Dynamics.** Comparison of network assortativity (circles) and the rate of change in cumulative number of UM employees (calculated over 5-year windows) from 2009 to 2023. Horizontal dashed lines indicate crossover from negative to positive values. Both metrics show a similar transition from negative to positive values around 2016, suggesting a connection between workforce changes and collaboration patterns

To explain the observed transition from disassortative to assortative mixing, we propose a resource-dependent collaboration framework grounded in network science literature. This framework assumes that researchers adapt their collaboration strategies based on available resources (Tomasello et al., 2017), and that these adaptations emerge from individual decisions rather than central coordination.

During resource-constrained periods (2009-2013), the academic system naturally adopts a mentorship-oriented structure where established researchers collaborate primarily with early-career researchers. This creates hierarchical, disassortative networks with distinct core-periphery structures (Csermely et al., 2013), efficiently distributing limited resources while facilitating knowledge transfer. As resources become more abundant (post-2016), the system shifts toward strategic clustering where well-connected researchers increasingly collaborate with equally established peers to maximize productivity and impact. This creates more assortative networks characterized by stronger interconnected clusters.

This framework aligns remarkably well with our empirical findings. It explains not only the synchronized transition of assortativity and employment rates from negative to positive around 2016 (Figure 12), but also accounts for the increasing clustering coefficient and enhanced network resilience to targeted attacks. The correlation between changing resource conditions and network reorganization suggests a causal relationship rather than coincidence—the resource environment fundamentally shapes how researchers form collaborations, leading to structural transitions in the co-authorship network (Uzzi & Spiro, 2005). This understanding offers valuable insights for research policy, suggesting that funding patterns and employment stability directly influence not only research output quantity but also the underlying structure of knowledge production.

Intriguingly, Figure 12 shows that this shift in assortativity closely parallels the rate of change in cumulative UM employment, with both quantities transitioning from negative to positive values around 2016.

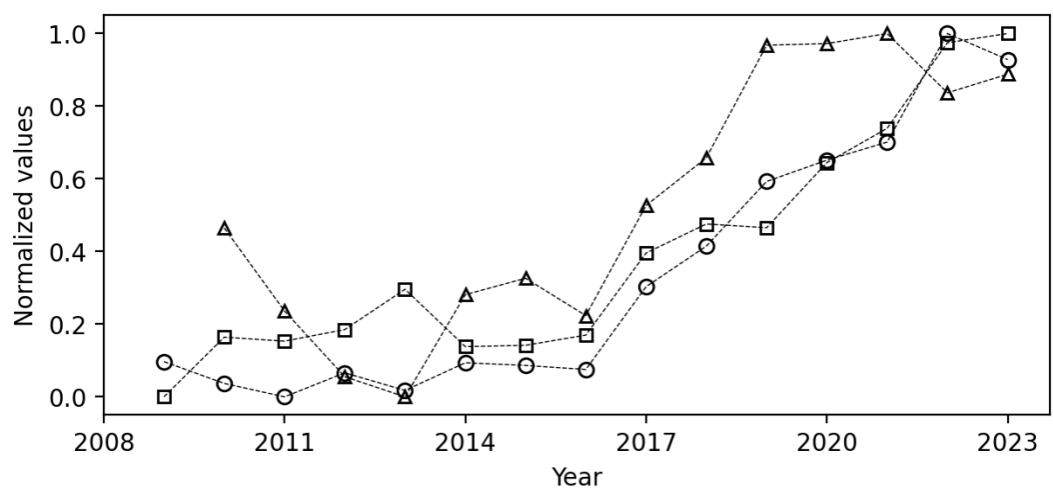

**Figure 13. Synchronized Evolution of Network Properties and Employment.** Normalized values of clustering coefficient, assortativity, and employee rate from 2009 to 2023, illustrating the remarkably similar temporal evolution of all three quantities. This visualization emphasizes the strong correlation between network structural properties and institutional workforce dynamics, suggesting interconnected mechanisms driving these changes.

Figure 13 further illustrates the parallel evolution of normalized assortativity, clustering coefficient, and employee rate, suggesting a deep interrelationship between network structure and workforce dynamics.

The comprehensive analysis of publication patterns, author dynamics, and network structure at the University of Maribor reveals a complex adaptive system that responded to economic pressures and policy changes in multifaceted ways. When examining these results collectively, a more nuanced picture emerges beyond what individual metrics might suggest. The divergence between employment figures and unique author counts during 2009-2013 demonstrates the system's adaptability—despite resource constraints, research output not only continued but accelerated. This resilience manifests differently across time scales, with high short-term volatility (annual Jaccard indices of 0.32-0.38 and churn rates of 40-50%) contrasting sharply with substantial mid-term stability (five-year Jaccard indices of 0.75-0.82 and churn rates of 7-11%). The survival analysis further illuminates this pattern, showing rapid initial attrition followed by remarkable persistence among established researchers. These temporal dynamics parallel structural changes in the co-authorship network, which has evolved toward greater resilience against targeted disruptions over the study period. The striking transition from dissassortative to assortative mixing around 2016 represents a fundamental reorganization in collaboration patterns, coinciding with recovery in employment growth and increased clustering. This structural shift appears closely coupled with institutional employment dynamics, as evidenced by the parallel trajectories of assortativity and employment rate changes. The remarkable synchronicity between network properties and workforce metrics suggests that the observed changes represent not merely statistical artifacts but profound transformations in how research is conducted and knowledge is produced at the university. Together, these findings illustrate a research ecosystem that maintained productivity through changing conditions by adapting both its composition and its interconnection patterns—reconfiguring itself while preserving core functions despite external pressures.

## 4. Discussion

Our findings provide a multifaceted perspective on the resilience and volatility of academic publishing at the University of Maribor between 2004 and 2023. We have demonstrated that the relationship between employed researchers and publishing authors is, in reality, a complex interplay influenced by economic conditions, funding models, evaluation policies, and the evolving structure of research collaboration (Abramo et al., 2017; Auranen & Nieminen, 2010).

The most striking initial observation is the divergence between the number of employed researchers and the number of unique authors during the economic recession (2009-2013). While employment figures declined sharply, mandatory retirement was also carried out at UM in accordance with national savings legislation, the number of unique authors continued to rise steadily. This phenomenon highlights the significant role of factors beyond direct, full-time employment in driving scientific output. Several interconnected factors likely contribute to this divergence.

The interplay between our various analytical approaches reveals a research ecosystem characterized by both fragility and remarkable resilience. When viewed holistically, the employment trends, authorship patterns, survival analyses, and network properties tell a coherent story of adaptation to changing circumstances. The system responded to economic pressures not merely by contracting but by reorganizing—maintaining and even increasing output through altered collaboration patterns rather than through workforce expansion. This adaptation manifested as a dual dynamic of short-term fluidity coupled with mid-term stability. The high annual turnover of authors, which might initially appear concerning, actually represents a healthy mechanism allowing the system to incorporate new contributors—particularly doctoral students and project-based researchers—while maintaining a stable core of experienced researchers who provide continuity. This core-periphery structure is further reflected in the survival analysis, where the initial steep decline in publishing activity gives way to a persistent plateau. The evolution of network properties adds crucial context to these patterns. The increasing resilience to targeted attacks suggests that the system has developed more distributed pathways of collaboration, reducing dependence on individual key researchers. Most significantly, the transition from dissassortative to assortative mixing around 2016 marks a profound shift in how researchers collaborate—moving from a model where highly-connected researchers primarily worked with less-connected ones to one where highly-connected researchers increasingly collaborate with each other. This structural transformation coincided with broader institutional recovery from austerity measures and likely reflects both strategic decisions in research management and organic responses to evolving incentive structures. The closely correlated trajectories of assortativity, clustering, and employment rates suggest that these changes represent a systemic reorganization rather than isolated phenomena. This comprehensive view underscores the inadequacy of evaluating research performance through simple metrics or short-term windows, highlighting instead the importance of understanding the dynamic interrelationships between workforce composition, collaboration patterns, and network structure in sustaining productive academic environments through changing conditions.

First, the increasing reliance on project-based funding creates a system where researchers are often employed on short-term contracts tied to specific projects (Auranen & Nieminen, 2010; Hicks, 2012). These researchers, while not counted as "permanent" employees, are highly motivated to publish to secure future funding and advance their careers. Second, the expansion of doctoral and postdoctoral programs has created a large pool of early-career researchers who contribute significantly to publication output, often as co-authors on papers led by more senior faculty (Kyvik & Aksnes, 2015). Third, the growing emphasis on international collaboration, incentivized by metric-based evaluation and funding criteria, expands the network of researchers contributing to UM-affiliated publications. Finally, the pervasive influence of bibliometric indicators in research evaluation creates strong pressure to publish (Franzoni et al., 2011), potentially increasing the number of individuals who engage in publishing, even if their primary affiliation is not with UM. In 2012, the internal Criteria for Elections to the Titles of Higher Education Teachers and Higher Education Associates of the UM will also include the requirement to demonstrate the international reputation of the scientific and professional work of researchers.

Our analysis of author overlap reveals a crucial distinction between short-term volatility and mid-term stability. The high annual churn rates (40-50%) and low annual Jaccard indices (0.32-0.38) suggest significant turnover in the annual author population. However, the 5-year analysis shows much higher Jaccard indices (0.75-0.82) and lower churn rates (8-10%), demonstrating that a substantial "core" group of researchers remains active over longer periods. This discrepancy highlights the prevalence of intermittent publishing (Kwiek & Szymula, 2024). Researchers may not publish every year but often reappear as authors within a 5-year window. The survival analyses corroborate this pattern, showing rapid initial decline in author activity followed by long-term persistence for a subset of researchers.

The network analysis adds another crucial dimension to our understanding. The UM co-authorship network has become increasingly resilient to targeted attacks over time, and the pattern of connections has changed in ways that enhance robustness. Most significantly, we observed a shift from a dissassortative to an assortative network structure around 2016, suggesting a fundamental change in collaboration patterns (Newman, 2002).

This shift could be driven by several factors: an increased focus on high-impact research incentivized by metric-based evaluation (Franzoni et al., 2011; Hicks, 2012); the emergence of strong, stable research groups creating a "rich-club" phenomenon (Colizza et al., 2006); changes in the funding landscape favoring larger collaborative projects; or strategic hiring decisions bringing in senior researchers with large existing networks.

The increasing assortativity and clustering could have complex effects on network resilience (Ahuja, 2000). While a more interconnected network might be more robust overall, a highly assortative structure could create vulnerabilities to the targeted removal of key individuals within dominant clusters. The close correspondence between the assortativity shift and changes in employment patterns suggests these structural changes are connected to broader institutional dynamics, possibly reflecting strategic decisions in hiring, changes in institutional priorities, or shifts in the research landscape (Fortunato et al., 2018).

Our findings have several important implications for research policy and university management. First, universities should recognize and support the "core" group of long-term researchers who provide stability, mentorship, and institutional knowledge (Kyvik & Aksnes, 2015; Levin & Stephan, 1991). Second, while some churn is inherent in the modern research environment, excessive turnover can be detrimental to institutional memory and mentoring capacity (Kwiek & Szymula, 2024). Third, metric-based evaluation should be implemented thoughtfully to avoid unintended consequences that might dilute focus on long-term, impactful research (Franzoni et al., 2011; Hicks, 2012). Fourth, universities should foster diverse collaborations to enhance network resilience (Moody, 2004; Uzzi & Spiro, 2005). Finally, multi-year perspectives should be adopted when evaluating research performance to account for fluctuating publication patterns (Hicks, 2012; Woelert & McKenzie, 2018).

This study has limitations including potential name ambiguity issues despite using unique author identifiers, focus on a single university (though with some comparison to the University of Ljubljana), challenges in establishing causality, lack of qualitative data, and constraints of the OpenAlex database. Future research could explore the impact of specific funding programs on collaboration patterns, the relationship between research quality and network structure, demographic factors in shaping research careers and network positions, and the long-term effects of the COVID-19 pandemic.

## 5. Conclusions

This study investigated the dynamics of academic publishing at the University of Maribor from 2004 to 2023, uncovering a complex interplay of factors shaping the resilience and volatility of the university's research ecosystem. Our findings challenge simplistic interpretations based solely on annual metrics and highlight the importance of considering longer-term trends and underlying network structure (Fortunato et al., 2018).

Our core findings reveal a multi-layered reality: First, despite declining employment during economic downturns, the number of unique authors continued to grow steadily (Abramo et al., 2017; Auranen & Nieminen, 2010), highlighting the adaptability of the academic system. Second, while annual metrics suggested high volatility (churn ~40-50%), the 5-year analysis revealed substantially greater stability (churn ~8-10%), demonstrating the existence of a core group of researchers who remain active over time (Kwiek & Szymula, 2024; Levin & Stephan, 1991). Third, the co-authorship network became increasingly resilient over time, with a more interconnected and robust structure. Fourth, we uncovered a profound structural transformation around 2016—a transition from negative to positive assortativity (Newman, 2002), indicating a shift from a system where highly connected researchers collaborated with less connected researchers to one where highly connected researchers increasingly collaborated with each other. This finding reflects a fundamental change in how research is conducted and aligns with global trends of increasing stratification in scientific collaboration (Fortunato et al., 2018).

These findings have profound implications for research evaluation and policy. Annual metrics can be misleading, overemphasizing short-term fluctuations and failing to capture underlying stability. High annual churn is not inherently negative in institutions with strong doctoral programs and project-based funding, though excessive turnover warrants attention. Network structure matters profoundly as an indicator of resilience and overall ecosystem health. The shift in assortativity presents a dilemma: while collaboration with established researchers can boost short-term metrics, over-reliance on this strategy could create a less inclusive and potentially more fragile system long-term (Moody, 2004; Uzzi & Spiro, 2005).

Based on our findings, we recommend strategic investment in core researchers through stable funding opportunities and career development programs; multi-year evaluation frameworks that acknowledge fluctuating research activity; promotion of network diversity through interdisciplinary projects and cross-department collaborations; careful monitoring of the effects of metric-based evaluation; and systematic data collection to track performance and inform policy decisions.

The University of Maribor, like many institutions worldwide, is navigating a complex environment shaped by global trends in academic publishing, funding models, and evaluation practices (Auranen & Nieminen, 2010; Geuna & Martin, 2003). Our study demonstrates that beneath apparent volatility lies a resilient core of researchers and an evolving network structure. However, the shift towards positive assortativity raises questions about long-term sustainability and potential vulnerabilities.

This research provides critical insights, highlighting that universities are complex, evolving networks rather than mere collections of individuals (Barabási et al., 2002; Newman, 2001). Understanding these dynamics is paramount for fostering a system that values collaboration, diversity, long-term research agendas, and contributions from researchers at all career stages (Fortunato et al., 2018; Uzzi & Spiro, 2005). A balanced approach—encouraging both strong research groups and broad, diverse collaborations—is essential for ensuring a vibrant, sustainable, and equitable research future.

**Funding:** DK received financial support from ARIS Slovenian Research and Innovation Agency, the research core funding program P3-0396. MTV received financial support from ARIS Slovenian Research and Innovation Agency, the research core funding program P2-0046.

**Code and data Availability:** The code to obtain and analyze the data used in this paper is available at https://github.com/deankorosak/resilience_volatility_publishing .